\documentclass[english]{article}
\usepackage{mathptmx}
\usepackage[T1]{fontenc}
\usepackage[latin9]{inputenc}
\usepackage{amsmath}
\usepackage{amsthm}
\usepackage{graphicx}
\usepackage[authoryear]{natbib}

\makeatletter
\renewcommand{\vec}[1]{{\bf #1}}

\makeatother

\usepackage{babel}
\begin{document}
\title{Unsupervised learning by a nonlinear network with Hebbian excitatory
and anti-Hebbian inhibitory neurons}
\author{H. Sebastian Seung\\
Neuroscience Institute and Computer Science Dept.\\
Princeton University, Princeton, NJ 08544\\
\texttt{sseung@princeton.edu}}
\maketitle
\begin{abstract}
This paper introduces a rate-based nonlinear neural network in which
excitatory ($E$) neurons receive feedforward excitation from sensory
($S$) neurons, and inhibit each other through disynaptic pathways
mediated by inhibitory ($I$) interneurons. Correlation-based plasticity
of disynaptic inhibition serves to incompletely decorrelate $E$ neuron
activity, pushing the $E$ neurons to learn distinct sensory features.
The plasticity equations additionally contain ``extra'' terms fostering
competition between excitatory synapses converging onto the same postsynaptic
neuron and inhibitory synapses diverging from the same presynaptic
neuron. The parameters of competition between $S\to E$ connections
can be adjusted to make learned features look more like ``parts''
or ``wholes.'' The parameters of competition between $I-E$ connections
can be adjusted to set the typical decorrelatedness and sparsity of
$E$ neuron activity. Numerical simulations of unsupervised learning
show that relatively few $I$ neurons can be sufficient for achieving
good decorrelation, and increasing the number of $I$ neurons makes
decorrelation more complete. Excitatory and inhibitory inputs to active
$E$ neurons are approximately balanced as a result of learning.
\end{abstract}
Network models combining Hebbian excitation and anti-Hebbian inhibition
have been explored previously by numerous researchers. The model of
\citet{foldiak1990forming} contained neurons that were recurrently
connected via all-to-all lateral inhibition, and received feedforward
connections from sensory afferents. Anti-Hebbian plasticity of inhibitory
connections forced the neurons to decorrelate their activities, thereby
enabling the Hebbian feedforward connections to learn distinct sensory
features. The term ``anti-Hebbian inhibition'' was used by \citet{foldiak1990forming}
to mean that correlated activity leads to strengthening of an inhibitory
connection (a negative number becomes more negative).

\begin{figure}
\begin{centering}
\includegraphics[width=0.6\textwidth]{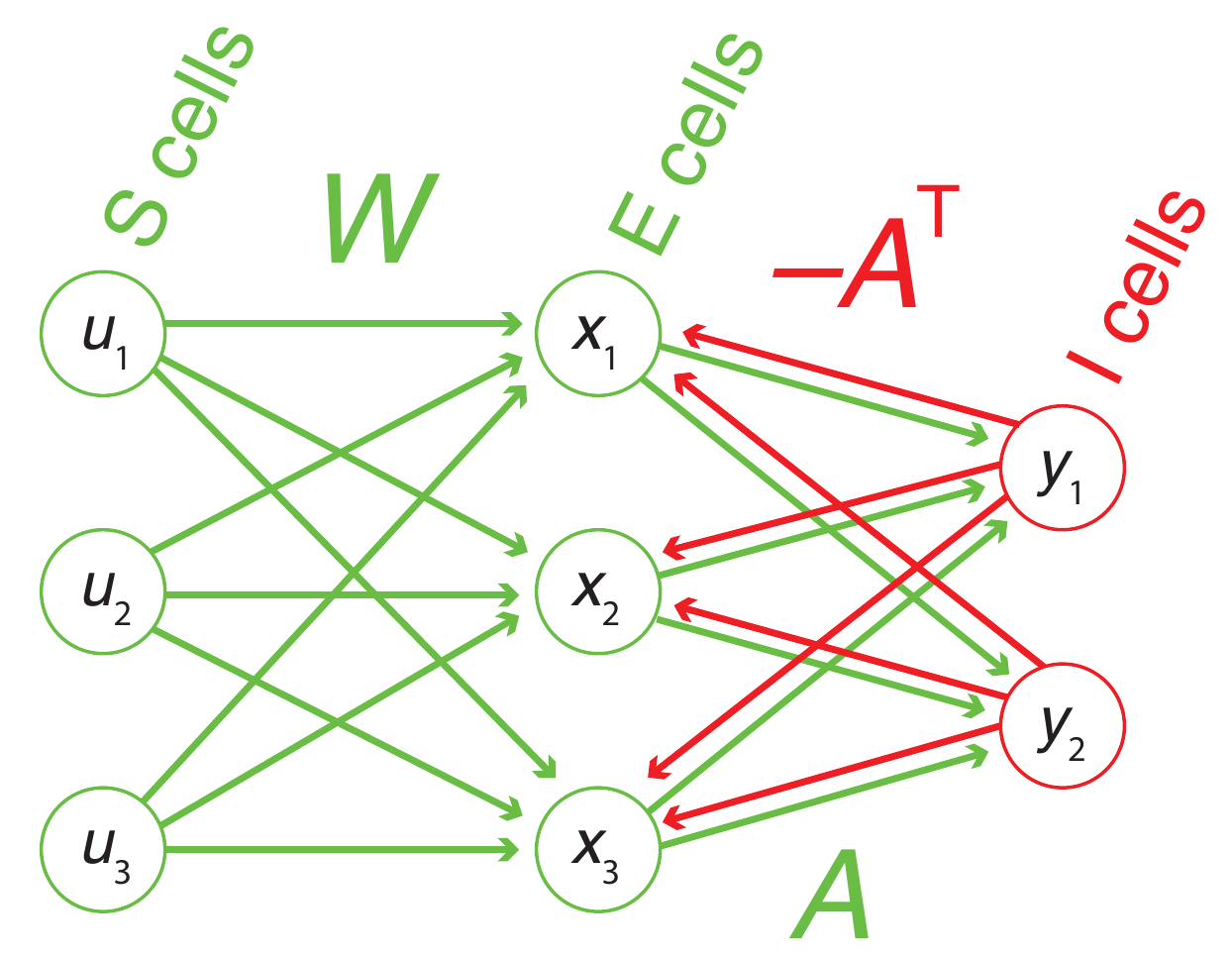}
\par\end{centering}
\caption{Excitatory-inhibitory net respecting Dale's Law. The $S\to E$ connection
from $u_{a}$ to $x_{i}$ has strength $W_{ia}$. The $E$ neurons
$x_{i}$ are reciprocally coupled with the $I$ neurons $y_{\alpha}$.
The $E\to I$ connection from $x_{i}$ to $y_{\alpha}$ has strength
$A_{\alpha i}$ and the $I\to E$ connection from $y_{\alpha}$ to
$x_{i}$ has strength $-A_{\alpha i}$. The green arrows indicate
excitatory connections ($W_{ia}\protect\geq0$, $A_{\alpha i}\protect\geq0$),
which change via Hebbian plasticity. The red arrows indicate inhibitory
connections (note minus sign), which change via anti-Hebbian plasticity.
The present model neglects $E\to E$ and $I\to I$ connections for
simplicity, in order to focus on the computational functions of $I-E$
connections and their plasticity.\label{fig:NetworkArchitectureDalesLaw}}
\end{figure}

\citet{king2013inhibitory} studied a model containing excitatory
($E$) neurons reciprocally connected with inhibitory ($I$) neurons,
and also receiving connections from sensory ($S$) afferents (Fig.
\ref{fig:NetworkArchitectureDalesLaw}). The $S\to E$ connections
were Hebbian, as in the \citet{foldiak1990forming} model. The $E$
neurons did not inhibit each other directly, but indirectly through
disynaptic $E\to I\to E$ pathways mediated by $I$ neurons. This
network motif will be called ``disynaptic recurrent inhibition,''
or just ``disynaptic inhibition.'' The $E\to I$ connections were
modified by Hebbian plasticity and the $I\to E$ connections by anti-Hebbian
plasticity. \citet{king2013inhibitory} showed that the $I$ neurons
could force the $E$ neurons to decorrelate their activities, enabling
Hebbian plasticity of feedforward connections to learn distinct sensory
features much as in the original \citet{foldiak1990forming} model.

This paper introduces a rate-based variant of the \citet{king2013inhibitory}
model, which used spiking integrate-and-fire model neurons for neurobiological
realism. A rate-based model is more amenable to mathematical understanding.
A rate-based model is also potentially useful for machine learning
applications, as it does not require the computational overhead of
simulating a spiking network. The novelty of the present model relative
to previous rate-based models is that it is (1) nonlinear, and (2)
separated into distinct $E$ and $I$ populations that respect Dale\textquoteright s
Law.\footnote{Here Dale's Law means that the outgoing synapses of a neuron are typically
either all excitatory or all inhibitory. The original formulation
of Dale's Law is that a neuron secretes a single neurotransmitter
at all of its outgoing synapses. The formulations are equivalent if
a single neurotransmitter has uniformly excitatory or uniformly inhibitory
influences on all postsynaptic neurons. There are known exceptions
in which a neuron secretes a single neurotransmitter that excites
some postsynaptic neurons and inhibits others, such as photoreceptor
synapses onto retinal bipolar cells \citep{euler2014retinal}, or
when a neuron secretes more than one neurotransmitter \citep{vaaga2014dual}.
Dale's Law is also known as Dale's Principle.} Some previous rate-based models were nonlinear but did not contain
separate $E$ and $I$ populations \citep{foldiak1990forming,hu2014hebbian,pehlevan2014hebbian,seung2017correlation}.
Other previous rate-based models had linear neurons that performed
principal component analysis, and separated principal neurons and
interneurons without respecting Dale\textquoteright s Law \citep{plumbley1993efficient,fyfe1995introducing,pehlevan2015normative}.

A first payoff of mathematical tractability is that the plasticity
equation for $I-E$ connections can be ``derived'' from the \citet{foldiak1990forming}
model, or more precisely from the variant due to \citet{seung2017correlation}.
The derivation results in a Hebbian/anti-Hebbian plasticity equation
plus two ``extra'' terms fostering synaptic competition. Competitive
interactions between synapses sharing the same presynaptic or postsynaptic
neuron have long been postulated by models of cortical development,
and are supported by neurobiological evidence \citep{miller1996synaptic}.
Surprisingly, the derivation suggests that the parameters of $I-E$
synaptic competition set the typical decorrelatedness and sparseness
of $E$ activity.

The derivation regards disynaptic inhibition as an approximate matrix
factorization of all-to-all inhibition.\footnote{The ``derivation'' may not really be mathematically justifiable,
because it involves approximating a Lagrange multiplier. More precise
mathematical justifications are given in a companion paper \citep{seung2018two}.} Due to the approximation, decorrelation is expected to be incomplete,
with more $I$ neurons leading to more complete decorrelation. Numerical
simulations of unsupervised learning from MNIST images of handwritten
digits show that good decorrelation is achieved with only a modest
number of $I$ neurons, similar to what was already reported by \citet{king2013inhibitory}.
A novelty here is that good decorrelation is achievable (at least
in this example) without $I\to I$ connections, which are lacking
in the present model but exist in \citet{king2013inhibitory}.

The mathematical form of competition between $S\to E$ connections
is not fixed by the derivation, and for simplicity is chosen to be
the same as for $I-E$ connections. Competition between $S\to E$
connections determines the sparsity of connectivity, and therefore
whether learned features look more like ``parts'' or ``wholes.''

Synaptic competition is linear in the connection strenths; the only
nonlinearity is the nonnegativity constraint. This simplicity lends
itself to mathematical analysis. The ``winners'' of synaptic competition
are connections to neurons with the most strongly correlated activities;
other connections vanish. The number of surviving nonzero connections
depends on the numerical parameters of synaptic competition, as well
as the values of the activity correlations. Adjusting the parameters
makes competition more or less ``winner-take-all,'' resulting in
sparser or fuller connectivity, respectively.

Numerical simulations show that the network exhibits approximate excitatory-inhibitory
balance after learning, in the sense that active $E$ cells receive
excitatory input that only slightly exceeds inhibitory input. This
has an intriguing correspondence with the excitatory-inhibitory balance
observed in cortical circuits \citep{isaacson2011inhibition}. Previous
models of balanced networks have relied upon nonmodifiable random
or structured connectivity \citep{deneve2016efficient}, rather than
connections shaped by plasticity.

\section{Network model: description and ``derivation''}

In the network model, $E$ neurons receive excitatory input from $S$
neurons and inhibit each other through disynaptic pathways mediated
by $I$ neurons (Fig. \ref{fig:NetworkArchitectureDalesLaw}). The
plasticity equation for $I-E$ connections is ``derived'' from a
previous model in which neurons monosynaptically inhibit each other
through all-to-all connections \citep{foldiak1990forming,seung2017correlation}.
In the previous model, anti-Hebbian inhibition served to decorrelate
the activity of the neurons. In the present model, anti-Hebbian $I$
neurons serve to \emph{imperfectly} decorrelate the activities of
$E$ neurons.

\subsection{Disynaptic recurrent inhibition}

The network of Fig. \ref{fig:NetworkArchitectureDalesLaw} has the
activity dynamics, 
\begin{align}
x_{i} & :=\left[\left(1-dt\right)x_{i}+dt\:\lambda_{i}^{-1}\left(\sum_{a=1}^{n}W_{ia}u_{a}-\sum_{\alpha=1}^{r}y_{\alpha}A_{\alpha i}\right)\right]^{+}\label{eq:ExcitatoryDynamics}\\
y_{\alpha} & =\sum_{i=1}^{m}A_{\alpha i}x_{i}\label{eq:InhibitoryActivity}
\end{align}
Here $dt$ is a step size parameter, which can be set at a small constant
value or adjusted adaptively (Appendix \ref{sec:GradientProjection}).
With the constraint $A_{\alpha i}\geq0$, the $x\to y$ connections
are excitatory, while the $y\to x$ connections are inhibitory. The
$x$ and $y$ variables are the activities of excitatory ($E)$ and
inhibitory ($I$) neural populations, respectively.\footnote{The mnemonics eXcitatory and Ynhibitory can be used to remember that
$E$ neuron activities are $x_{i}$ while $I$ neuron activities are
$y_{\alpha}$. Inhibition has infinite speed according to Eq. (\ref{eq:InhibitoryActivity}).
Inhibition with finite speed could be implemented by some discrete
time approximation to $\tau dy_{\alpha}/dt+y_{\alpha}=\sum_{i}A_{\alpha i}x_{i}$.
For the continuous time case, convergence to a steady state can be
proven for sufficiently small $\tau$ \citep{seung1998minimax}. Similarly,
inhibition was faster than excitation in the model of \citet{king2013inhibitory}.} The $u$ variables are the activities of sensory ($S$) afferents.
The $u\to x$ connections are also excitatory, $W_{ia}\geq0$. The
numbers of $S$, $E$, and $I$ cells are $n$, $m$, and $r$, respectively.

The activation function $\left[z\right]^{+}=\max\left\{ z,0\right\} $
is half-wave rectification. If $x_{i}$ is initially negative, it
becomes nonnegative at the next time step and for all future times,
by Eq. (\ref{eq:ExcitatoryDynamics}). A rectification nonlinearity
could also be included in Eq. (\ref{eq:InhibitoryActivity}), but
would have no effect because both $A$ and $x$ are nonnegative.

After the activities converge to a steady state, update the connection
matrices via

\begin{align}
\Delta W_{ia} & \propto x_{i}u_{a}-\gamma W_{ia}-\kappa\sum_{b}W_{ib}\label{eq:UpdateW}\\
\Delta A_{\alpha j} & \propto y_{\alpha}x_{j}-\left(q^{2}-p^{2}\right)A_{\alpha j}-p^{2}\sum_{i}A_{\alpha i}\label{eq:UpdateA}
\end{align}
where $\gamma>0$, $\kappa>0$, and $q^{2}>p^{2}$ . After the updates
(\ref{eq:UpdateW}) and (\ref{eq:UpdateA}), any negative elements
of $W$ and $A$ are zeroed to maintain nonnegativity.\footnote{The spiking network model of \citet{king2013inhibitory} allowed the
nonspiking sensory inputs $u_{a}$ and the $S-E$ connections $W_{ia}$
to have arbitrary signs; the signs are constrained in the present
model. The plasticity equation for $W_{ia}$ was Hebbian plus weight
decay of the form proposed by \citet{oja1982simplified}. For other
classes of connections, \citet{king2013inhibitory} introduced a ``Correlation
Measuring'' rule, which was designed to make connection strength
proportional to the covariance of presynaptic and postsynaptic activity
at the stationary state of learning. The activities of both $E$ and
$I$ neurons were regulated by a homeostatic rule; in the present
model only the $E$ activities are regulated. \citet{king2013inhibitory}
used the homeostatic rule of \citet{foldiak1990forming} rather than
Eq. (\ref{eq:Homeostatic}) which was also used by \citet{seung2017correlation}.
\citet{king2013inhibitory} included $I\to I$ connections whereas
the present model omits them.}

The $\Delta W_{ia}$ update is said to be Hebbian, because positive
correlation between $x_{i}$ and $u_{a}$ causes $W_{ia}$ to become
more positive. Note that $A_{\alpha j}$ is the strength of the excitatory
$x_{i}\to y_{\alpha}$ connection, and $-A_{\alpha j}$ is the strength
of the inhibitory $y_{\alpha}\to x_{i}$ connection. The $\Delta A_{\alpha j}$
update is said to be Hebbian when it refers to the excitatory connection,
and anti-Hebbian when it refers to the inhibitory connection. The
term ``anti-Hebbian'' is used because positive correlation makes
the inhibitory connection more negative.

The connections between $E$ and $I$ neurons are equal and opposite,
or antisymmetric. The antisymmetry is a natural outcome of the fact
that $E\to I$ connections are Hebbian while $I\to E$ connections
are anti-Hebbian, so that the same update (\ref{eq:UpdateA}) applies
to both. Empirical evidence for approximate antisymmetry of $E\leftrightarrow I$
connections in mouse cortex has been reported by \citet{znamenskiy2018functional}.

The divisive factor $\lambda_{i}>0$ can be interpreted as the inverse
slope of the activation function, or a scale factor that divides the
connections converging onto the $i$th $E$ neuron. It is updated
via

\begin{equation}
\Delta\lambda_{i}\propto x_{i}^{2}-q^{2}\label{eq:Homeostatic}
\end{equation}
If squared activity $x_{i}^{2}$ is higher than the set point $q^{2}$,
then the update increases $\lambda_{i}$, thereby lowering activity
in the future. If squared activity $x_{i}^{2}$ is lower than the
set point $q^{2}$, then the update decreases $\lambda_{i}$, thereby
raising activity in the future. Therefore Eq. (\ref{eq:Homeostatic})
amounts to homeostatic regulation of activity so that $\langle x_{i}^{2}\rangle\approx q^{2}$.

\subsection{\label{subsec:All2All}``Derivation'' of $I-E$ plasticity equation}

Eqs. (\ref{eq:ExcitatoryDynamics}) and (\ref{eq:InhibitoryActivity})
can be combined into the dynamics

\begin{equation}
x_{i}:=\left[\left(1-dt\right)x_{i}+dt\,\lambda_{i}^{-1}\left(\sum_{a}W_{ia}u_{a}-\sum_{j}\left(A^{\top}A\right)_{ij}x_{j}\right)\right]^{+}\label{eq:ExcitatoryInhibitoryAll2All}
\end{equation}
This equation has all-to-all inhibitory connections between the $x$
neurons (which are now inhibitory rather than excitatory). Combined
with the plasticity equations (\ref{eq:UpdateW})-(\ref{eq:Homeostatic}),
Eq. (\ref{eq:ExcitatoryInhibitoryAll2All}) resembles a variant of
the \citet{foldiak1990forming} model due to \citet{seung2017correlation}.
The dynamics of activity in the previous model were\footnote{A subtlety is that the diagonal term $\left(A^{\top}A\right)_{ii}$
does not vanish, so that Eqs. (\ref{eq:ExcitatoryInhibitoryAll2All})
and (\ref{eq:NetworkDynamicsSeungZung}) are not exactly equivalent.
However, one can show that the Lyapunov functions of the activity
dynamics are equivalent with the identification $L=\Lambda+A^{\top}A$.}

\begin{equation}
x_{i}:=\left[(1-dt)x_{i}+dt\:L_{ii}^{-1}\left(\sum_{a}W_{ia}u_{a}-\sum_{j,j\neq i}L_{ij}x_{j}\right)\right]^{+}\label{eq:NetworkDynamicsSeungZung}
\end{equation}
The elements of the matrix $L$ were updated via\footnote{For $i\neq j$, Eq. (\ref{eq:SeungZungUpdate}) is exactly as in \citet{foldiak1990forming}.
For $i=j$, Eq. (\ref{eq:SeungZungUpdate}) is based on $x_{i}^{2}$
and divisively modifies Eq. (\ref{eq:NetworkDynamicsSeungZung}) while
the analogous equation in \citet{foldiak1990forming} is based on
$x_{i}$ and subtractively modifies the neural network dynamics. } 
\begin{equation}
\Delta L_{ij}\propto x_{i}x_{j}-D_{ij}\label{eq:SeungZungUpdate}
\end{equation}
where
\begin{equation}
D_{ij}=\begin{cases}
p^{2}, & i\neq j,\\
q^{2}, & i=j
\end{cases}\label{eq:DesiredCorrelations}
\end{equation}
\citet{seung2017correlation} showed formally that $L$ is a Lagrange
multiplier that enforces the constraint $\langle x_{i}x_{j}\rangle\leq D_{ij}$,
where $\langle\rangle$ denotes an average over stimuli, and that
Eq. (\ref{eq:SeungZungUpdate}) can be viewed as a gradient update
$\Delta L_{ij}\propto\partial S/\partial L_{ij}$ for some function
$S$. 

Now consider the parametrization $L=\Lambda+A^{\top}A$, meaning that
$L-\Lambda$ has a factorized form. Then a gradient update for $A$
follows from naive application of the chain rule,

\begin{align}
\Delta A_{\alpha j} & \propto\frac{\partial S}{\partial A_{\alpha j}}=\sum_{i}A_{\alpha i}\frac{\partial S}{\partial L_{ij}}=\sum_{i}A_{\alpha i}\left(x_{i}x_{j}-D_{ij}\right)\label{eq:ChainRuleDerivation}
\end{align}
Eq. (\ref{eq:UpdateA}) follows from Eqs. (\ref{eq:DesiredCorrelations})
and (\ref{eq:ChainRuleDerivation}). A gradient update for the diagonal
matrix $\Lambda$ similarly leads to Eq. (\ref{eq:Homeostatic}).

The preceding ``derivation'' regards the disynaptic inhibition of
the current model as an approximation to all-to-all inhibition. The
approximation has low rank if $I$ neurons are less numerous than
$E$ neurons. ``Derivation'' is in quotes because the parametrized
form for a Lagrange multiplier may not really be justifiable. Alternative
mathematical interpretations of the model are given in a companion
paper \citep{seung2018two}.

To summarize, the role of the plasticity equation (\ref{eq:UpdateA})
is to approximately enforce the constraint $\langle x_{i}x_{j}\rangle\leq D_{ij}$,
or incompletely decorrelate $E$ neuron activity. Decorrelation is
expected to become more complete with increasing number of $I$ neurons,
because the approximation $L=\Lambda+A^{\top}A$ should improve as
the rank of $A$ increases. In the following, completeness of decorrelation
will be assessed using numerical simulations.

\section{Numerical demonstration of decorrelation}

Numerical simulations of the network were done for sensory stimuli
drawn from the MNIST images of handwritten digits. The images were
normalized so that the minimum and maximum pixel values were 0 and
1, respectively. The network saw each of the 60,000 training images
once during learning.

The network contained 784 $S$ neurons, 64 $E$ neurons and $5$ $I$
neurons. The elements of $W$ were drawn from a uniform distribution,
and normalized so that $\sum_{a}W_{ia}=1$ for all $i$. The elements
of $A$ were drawn from a uniform distribution on the interval $[0,0.1]$.
The $\lambda_{i}$ were initialized at unity. The learning rate parameters
were 0.1 for $A$ and $\Lambda$ and $0.001$ for $W$. After each
$\lambda_{i}$ update via Eq. (\ref{eq:Homeostatic}), a bound constraint
$\lambda\geq\lambda_{min}$ was applied with $\lambda_{min}=0.01$.
This is not always necessary, but is helpful for avoiding numerical
instability in some cases.

The base parameter configuration used $\gamma/\kappa=5$ (specifically
$\kappa=0.01$, $\gamma=0.05$) as the synaptic competition parameters
in the $W$ update (\ref{eq:UpdateW}) and $p/q=1/3$ (specifically
$p=0.03$, $q=0.09$) in the $A$ update (\ref{eq:UpdateA}). Other
parameter configurations differed by altering one parameter of the
base configuration, so that $\gamma=0.5$ (Fig. \ref{fig:LearnedFeatures}),
$p=0.06$ (Figs. \ref{fig:SparsityVsTime}, \ref{fig:AllActivities},
\ref{fig:EECorrelationHistogram}, \ref{fig:SparserActivityFullerA}),
or $r=1,$10 (Fig. \ref{fig:SimilaritiesRank}). Note that the learned
representations (up to trivial rescalings) depend on $\kappa$, $\gamma$,
$p$, and $q$ only through the ratios $\kappa/\gamma$ and $p/q$.\footnote{Transforming a steady state of learning by doubling $W$ and $\Lambda$,
scaling $A$ and $y$ by $\sqrt{2}$, and holding $x$ fixed is a
steady state of learning for $\kappa$ and $\gamma$ halved. Transforming
a steady state of learning by doubling $x$, $y$, and $W$ while
holding $A$ and $\Lambda$ fixed yields a steady state of learning
for $p$ and $q$ doubled. This can be verified in an average velocity
approximation, according to which a steady state of learning satisfies
$\langle\Delta W_{ia}\rangle=0$ and $\langle\Delta A_{\alpha i}\rangle=0$.}

\subsection{The $E$ neurons learn sensory features}

\begin{figure}
\begin{centering}
\includegraphics[width=0.8\textwidth]{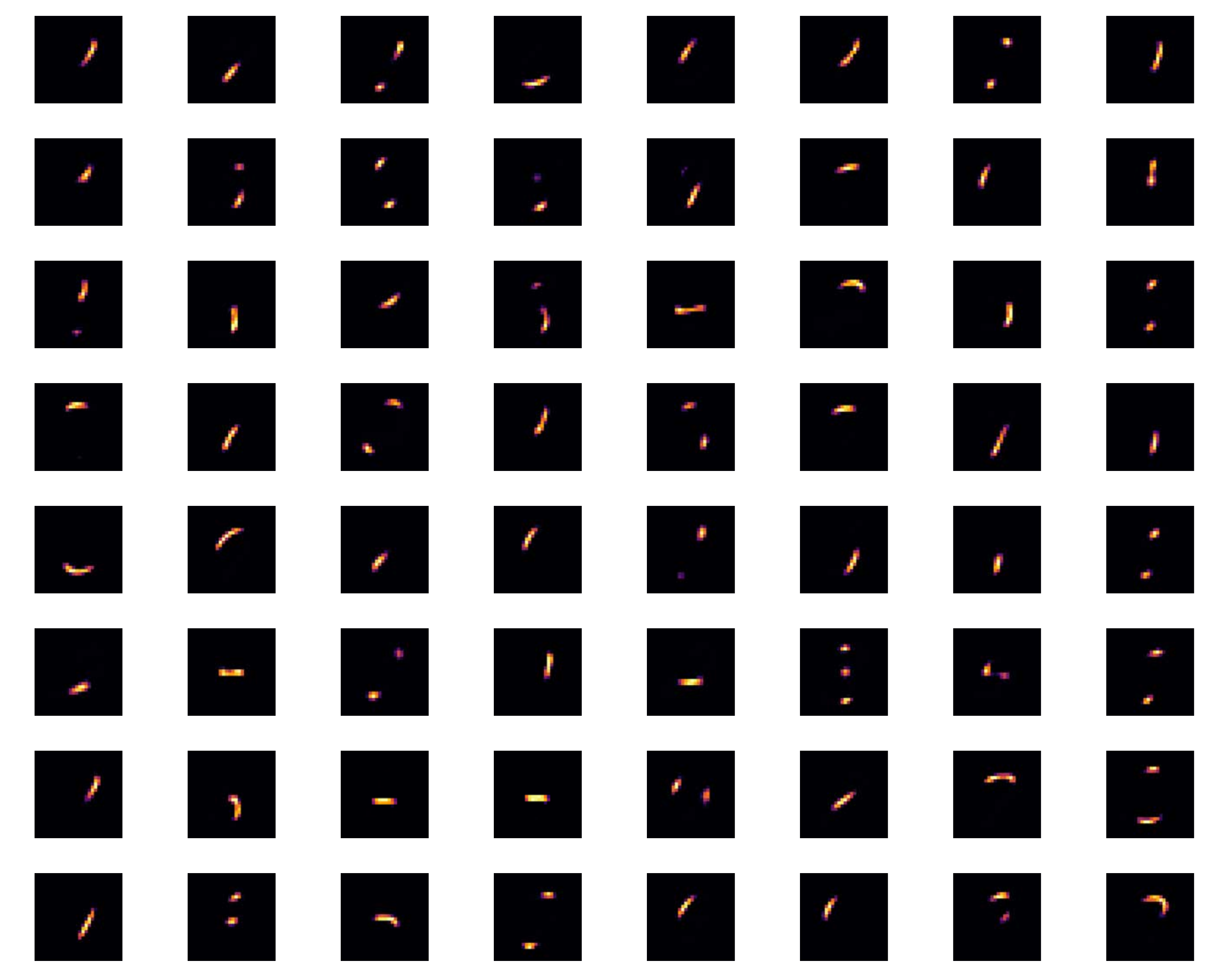}
\par\end{centering}
\vspace{1cm}

\begin{centering}
\includegraphics[width=0.8\textwidth]{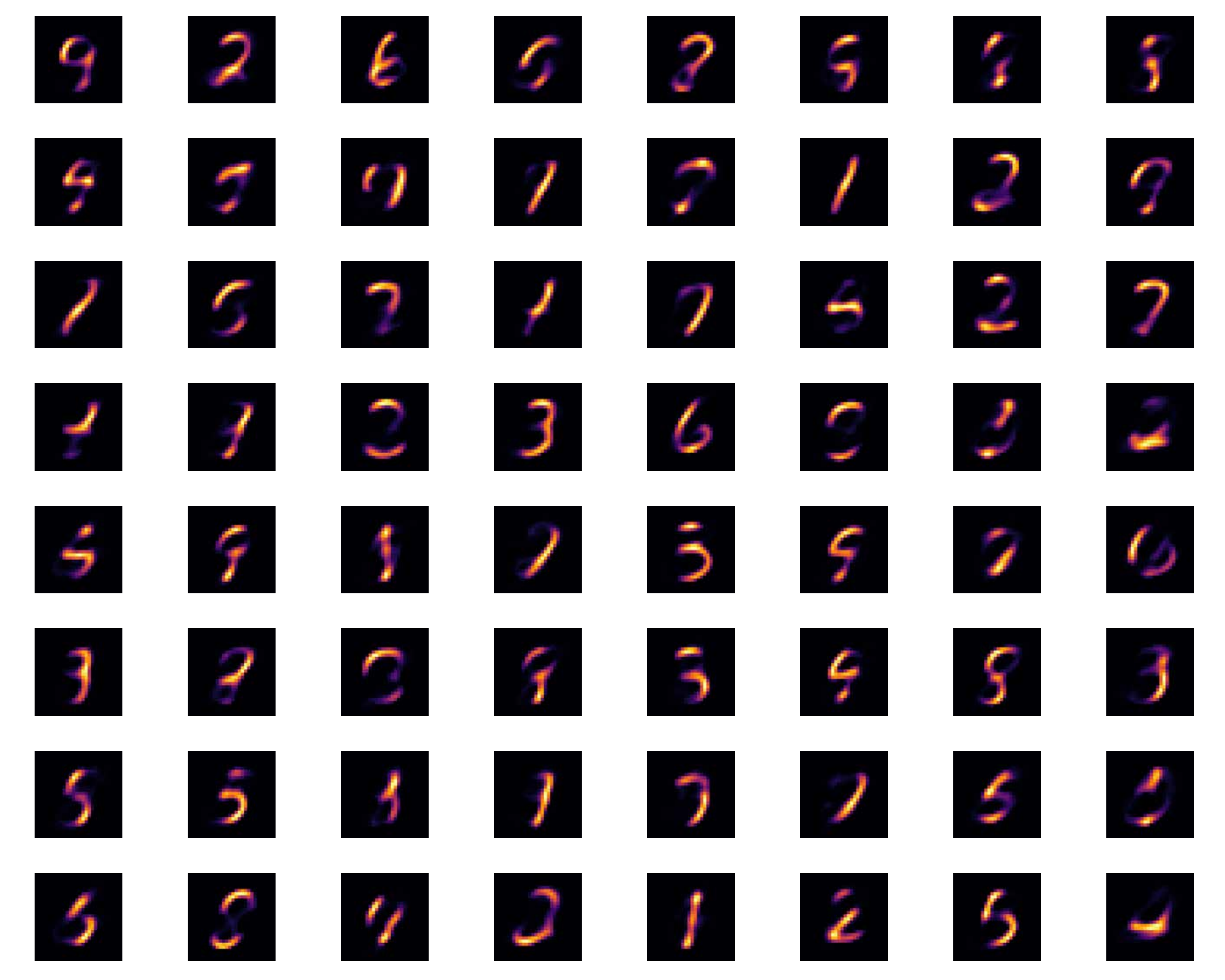}
\par\end{centering}
\caption{For each $E$ neuron, the convergent $S\to E$ connections constitute
a sensory feature learned from the stimuli. (a) Learned features for
$\gamma/\kappa=5$ are more sparse, resembling character ``strokes.''
(b) Learned features for $\gamma/\kappa=50$ are more full, looking
almost like entire digits.\label{fig:LearnedFeatures}}
\end{figure}
For each of the 64 $E$ neurons, the set of convergent $S\to E$ connections
can be viewed as an image (Fig. \ref{fig:LearnedFeatures}). The images
will be called ``sensory features,'' or simply ``features.'' For
the base parameter configuration ($\gamma/\kappa=5$), the features
are more sparse and resemble character ``strokes.'' For $\gamma/\kappa=50$,
the features are more full, and look almost like entire digits. By
controlling sparsity of $S\to E$ connections, the ratio $\gamma/\kappa$
effectively determines whether $E$ neurons learn ``parts'' or ``wholes.''

\subsection{\emph{E} neurons are more selective than \emph{I} neurons}

\begin{figure}
\begin{centering}
\includegraphics[width=0.7\textwidth]{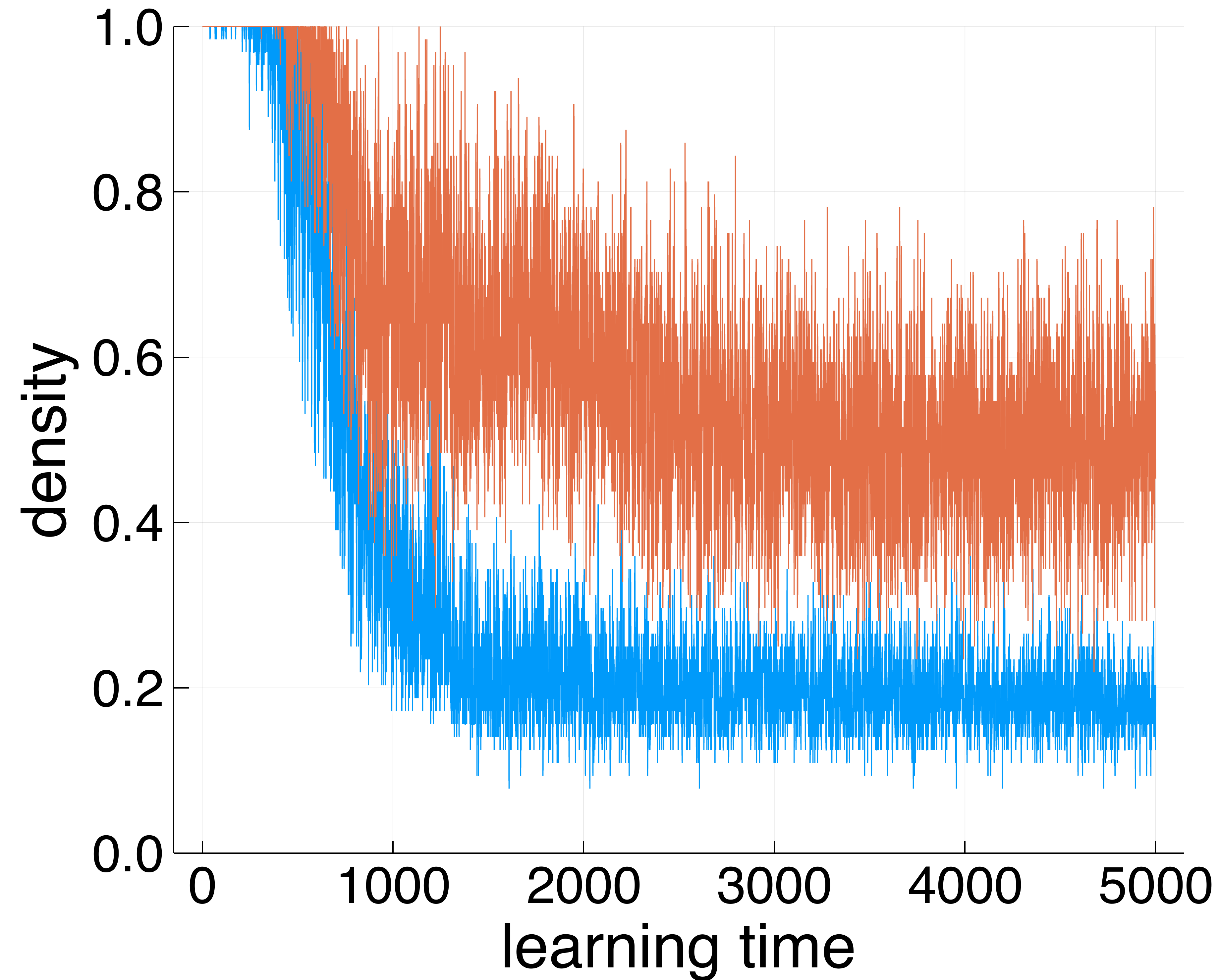}
\par\end{centering}
\caption{$E$ neuron activity starts out full, and sparsens as learning proceeds.
Activity ends up sparser for $p/q=1/3$ (blue) than for $p/q=2/3$
(red).\label{fig:SparsityVsTime}}
\end{figure}

\begin{figure}
\begin{centering}
\includegraphics[width=1\textwidth]{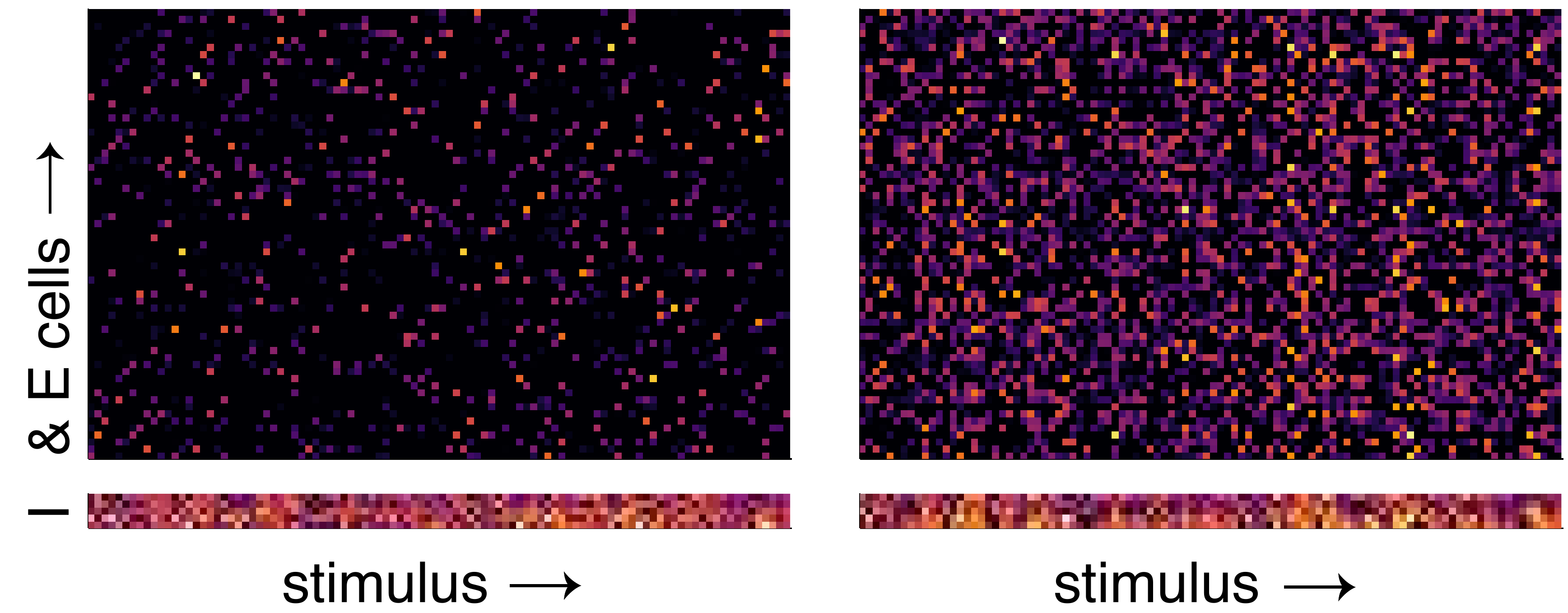}
\par\end{centering}
\caption{$E$ cells (top) are more sparsely active than $I$ cells (bottom),
which are fully active. $E$ cell activity is more sparse for $p/q=1/3$
(left) and more full for $p/q=2/3$ (right).\label{fig:AllActivities}}
\end{figure}

At the beginning of learning, all $E$ neurons are active when the
network dynamics (\ref{eq:ExcitatoryDynamics}) converges to a steady
state. As learning proceeds, some neurons become inactive at the steady
state (Fig. \ref{fig:SparsityVsTime}). Once learning has converged,
$E$ activity is sparse at the steady state, while all $I$ neurons
are active for all stimuli (Fig. \ref{fig:AllActivities}). The activity
of $E$ neurons is more sparse for smaller values of $p/q$ (Figs.
\ref{fig:SparsityVsTime} and \ref{fig:AllActivities}).

\begin{figure}
\begin{centering}
\includegraphics[width=0.7\textwidth]{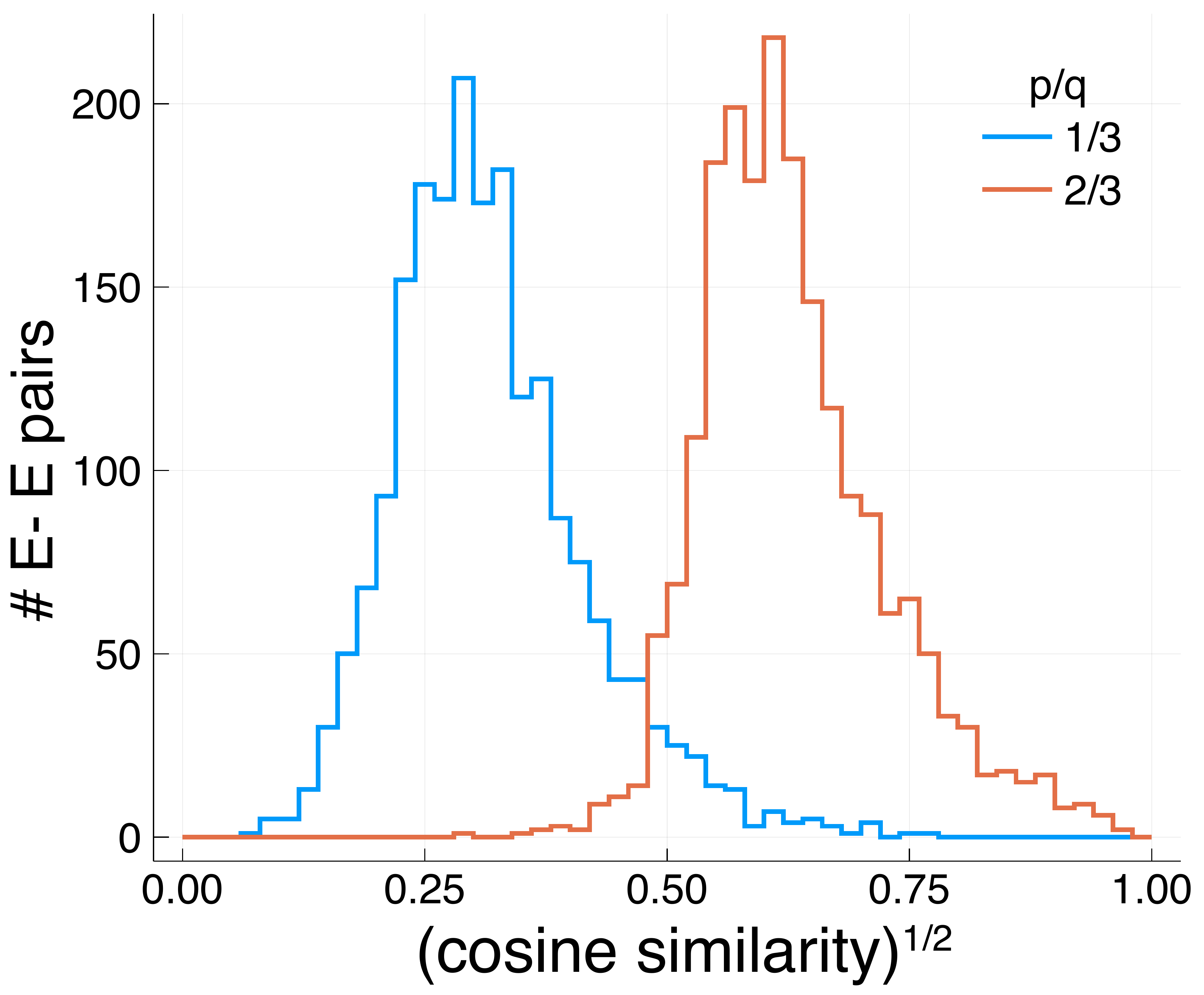}
\par\end{centering}
\caption{The ratio $p/q$ controls the degree of decorrelation. Histograms
of the square root of the cosine similarity of $E-E$ pairs for $p/q=1/3$
(blue) and $p/q=2/3$ (red). The histograms are peaked near $p/q$,
roughly in accord with Eq. (\ref{eq:CosineSimilarityBound}).\label{fig:EECorrelationHistogram}}
\end{figure}

This behavior can be understood from the idea, introduced in Section
\ref{subsec:All2All}, that anti-Hebbian inhibition enforces the constraint
$\langle x_{i}x_{j}\rangle\leq D_{ij}$. If equality holds for $i=j$,
the constraint takes the form
\begin{equation}
\frac{\langle x_{i}x_{j}\rangle}{\sqrt{\langle x_{i}^{2}\rangle\langle x_{j}^{2}\rangle}}\leq\frac{p^{2}}{q^{2}},\label{eq:CosineSimilarityBound}
\end{equation}
where the left hand side is the cosine similarity of the activities
of $E$ neurons $i$ and $j$. In other words, anti-Hebbian inhibition
encourages the activities of $E$ neurons to be dissimilar or decorrelated
if $p/q$ is small (Fig. \ref{fig:EECorrelationHistogram}). Because
the activities are nonnegative, decorrelation leads to sparsity.

\subsection{Decorrelation is incomplete}

\begin{figure}
\begin{centering}
\includegraphics[width=0.7\textwidth]{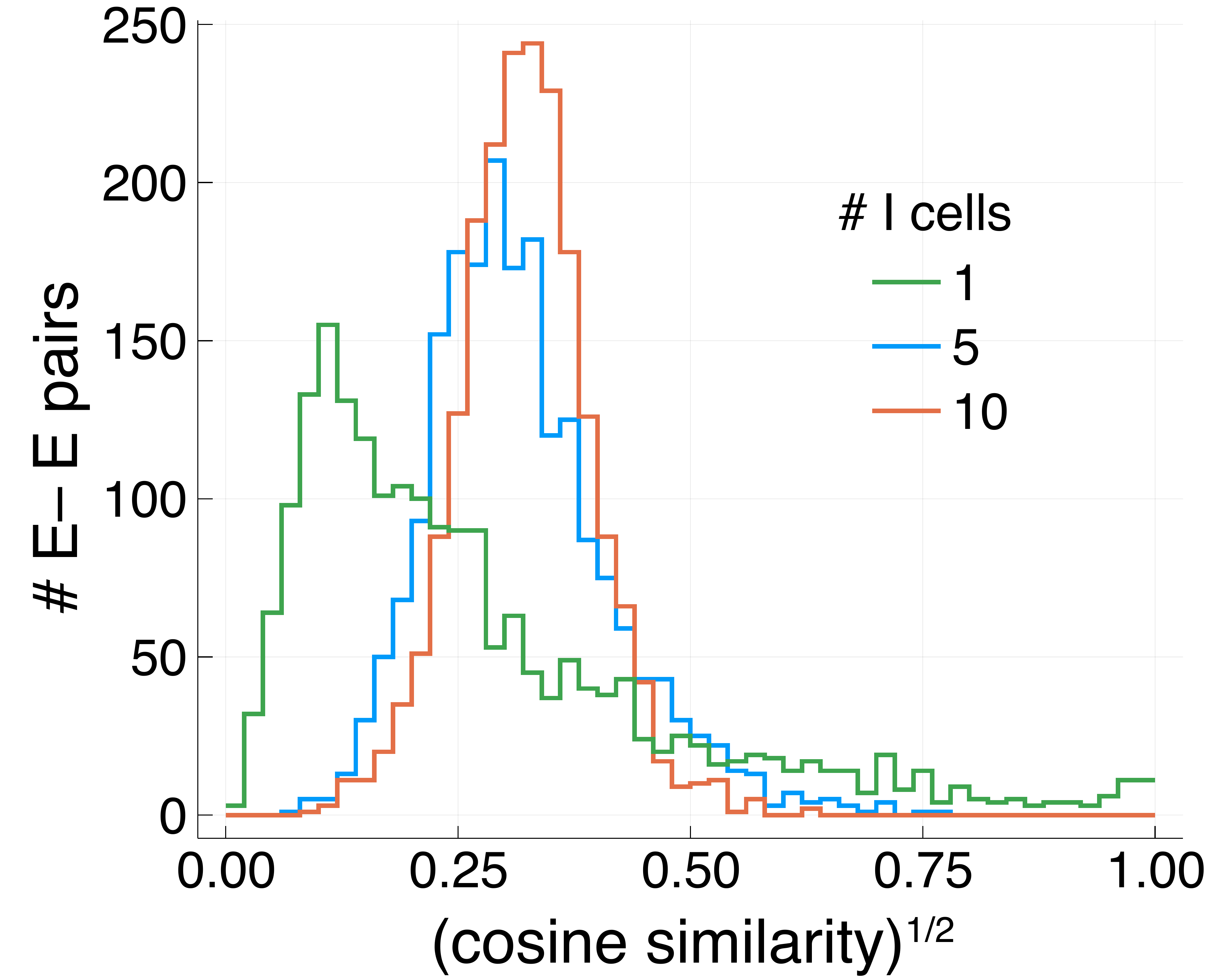}
\par\end{centering}
\caption{Decorrelation of $E$ neuron activity becomes more complete with more
$I$ neurons. Histograms are of the square root of cosine similarity
with $p/q=1/3$. For a single $I$ neuron (green), the histogram peaks
well below $p/q$, and has a long tail to the right. For 5 $I$ neurons
(blue), the histogram is peaked near $p/q$ and the tail to the right
is reduced. For 10 $I$ neurons (red) the tail to the right is further
reduced.\label{fig:SimilaritiesRank}}

\end{figure}

By the arguments of Section \ref{subsec:All2All}, we expect that
decorrelation should be less complete when $I$ cells are less numerous.
Indeed, reducing to a single $I$ neuron results in a long tail of
highly correlated $E$ cell pairs (Fig. \ref{fig:SimilaritiesRank}).
The tail is reduced in the base parameter configuration of $5$ $I$
neurons, and further reduced when the number of $I$ neurons is increased
to 10 (Fig. \ref{fig:SimilaritiesRank}).\footnote{One might expect that complete decorrelation is guaranteed at $r=n$
(full rank). However this is not necessarily true as $A$ is constrained
to be nonnegative.}

\section{Synaptic competition}

The first terms of the plasticity equations (\ref{eq:UpdateW}) and
(\ref{eq:UpdateA}) are nonnegative, since all neural activities are
nonnegative. Therefore the connection strengths could increase without
bound, were it not for the ``extra'' terms in the plasticity equations.
In the plasticity equation (\ref{eq:UpdateW}) for $\Delta W_{ia}$,
the ``weight decay'' term $-\gamma W_{ia}$ is said to be homosynaptic,
since the change in the connection depends on the strength of the
same connection. The other term $-\kappa\sum_{b}W_{ib}$ is heterosynaptic,
since through it $\Delta W_{ia}$ depends on all other connections
converging onto the $i$th $E$ cell.\footnote{In models of cortical development, it is more common to constrain
the sums in Eqs. (\ref{eq:UpdateW}) and (\ref{eq:UpdateA}), so that
for example $\sum_{b}W_{ib}=\rho$ \citep{von1973self,miller1994role}.
This has the biological interpretation that the synapses are competing
for fixed amount of resources. The constraint can be implemented in
a ``soft'' way by substituting $\sum_{b}W_{ib}-\rho$ for $\sum_{b}W_{ib}$
in Eq. (\ref{eq:UpdateW}), as was done by \citet{seung2017correlation}.
Here the constraint is eliminated altogether to reduce the number
of parameters and simplify the model.}

The roles of the two terms can be explained with a financial analogy.
The weight decay term $-\gamma W_{ia}$ is larger for stronger connections,
and therefore acts like a ``flat tax.'' The heterosynaptic term
$-\kappa\sum_{a}W_{ia}$ is the same for all connections, large or
small, converging onto neuron $i$. This amounts to a ``regressive
tax'' on convergent connections. By itself the regressive tax would
lead to ``winner-take-all'': all connections would vanish except
for a single winner. The flat tax is more egalitarian: adding it prevents
a single winner from dominating.

An analogous combination of homosynaptic and heterosynaptic terms
is also found in the plasticity equation (\ref{eq:UpdateA}) for $\Delta A_{\alpha i}$.
In this case the heterosynaptic term fosters competition between inhibitory
connections diverging from the $\alpha$th $I$ neuron and excitatory
connections converging onto the $\alpha$th $I$ neuron.
\begin{figure}
\begin{centering}
\includegraphics[width=1\textwidth]{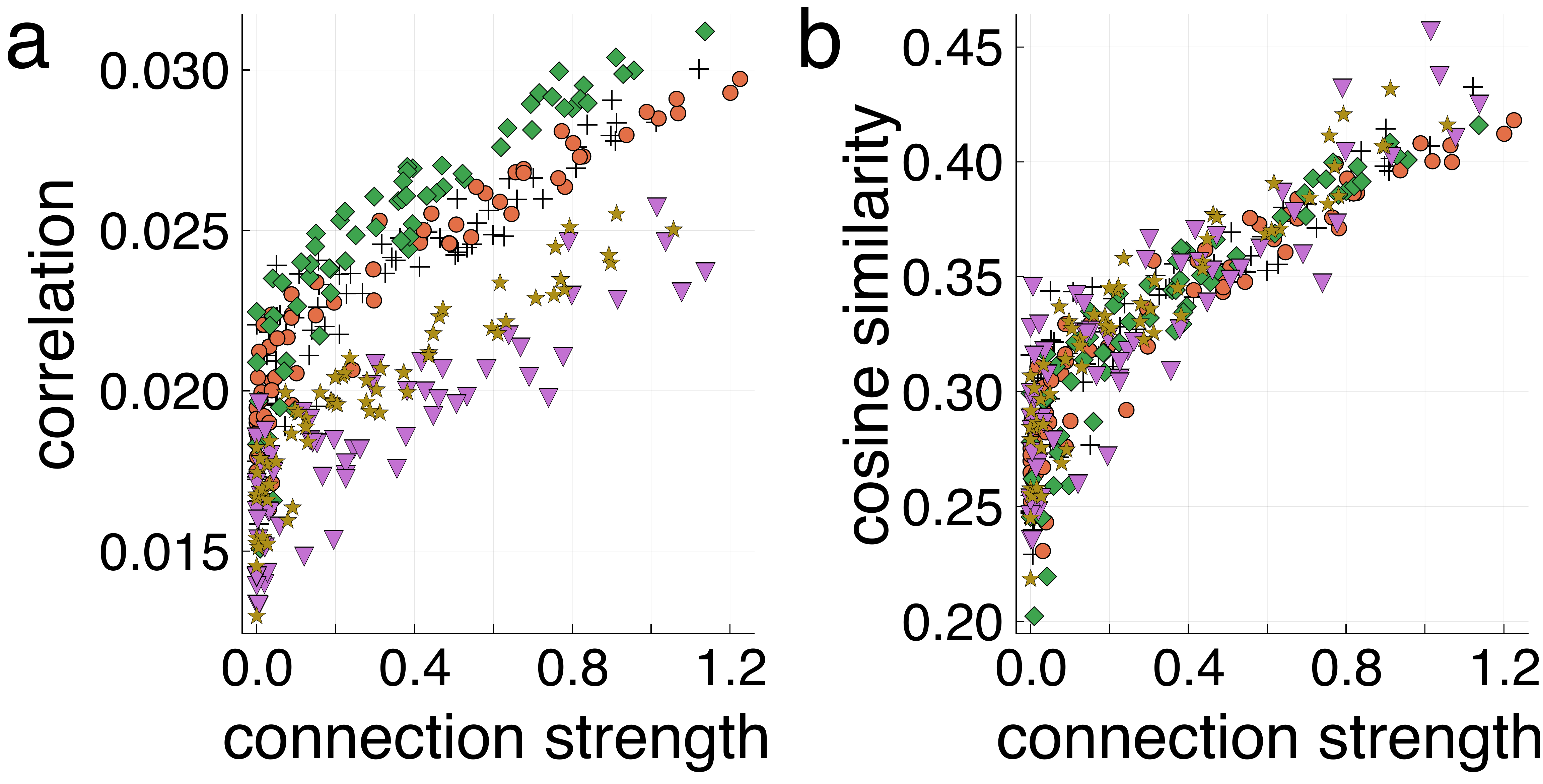}
\par\end{centering}
\caption{$I-E$ connection strength is linearly related to correlation or response
similarity. The connections of each $I$ neuron are distinguished
by color and marker type. (a) Correlation $\langle y_{\alpha}x_{i}\rangle$
versus connection strength $A_{\alpha i}$ has the same slope but
different intercept for each $I$ neuron, consistent with Eq. (\ref{eq:CorrelationConnectionDisynapticInhibition}).
(b) The relation of cosine similarity to connection strength varies
less across $I$ neurons.\label{fig:EICorrelationConnection}}
\end{figure}

Competition between synapses is simple enough for analytical treatment.
Eq. (\ref{eq:UpdateA}) and nonnegativity of $A$ imply that

\begin{equation}
\left(q^{2}-p^{2}\right)A_{\alpha j}\approx\left[\langle y_{\alpha}x_{j}\rangle-p^{2}\sum_{i}A_{\alpha i}\right]^{+}\label{eq:CorrelationConnectionDisynapticInhibition}
\end{equation}
at a stationary state of learning. Numerical simulations (Fig. \ref{fig:EICorrelationConnection})
agree well with Eq. (\ref{eq:CorrelationConnectionDisynapticInhibition}).
For each $I$ neuron $\alpha$, the top $k$ connections are linearly
related to the top $k$ correlations, while the rest of the connections
vanish. The number $k$ of surviving connections varies from neuron
to neuron. It depends on the ratio $p/q$ as well as the values of
the activity correlations through inequalities derived in Appendix
\ref{sec:NumberSurvivingConnections}. Simple statements can be made
for limiting cases of $p/q$. The competition is ``winner-take-all''
as $p/q\to1$, with a single nonzero connection surviving for each
$I$ neuron. As $p/q\to0$, on the other hand, all connections survive.

\begin{figure}
\begin{centering}
\includegraphics[width=0.7\textwidth]{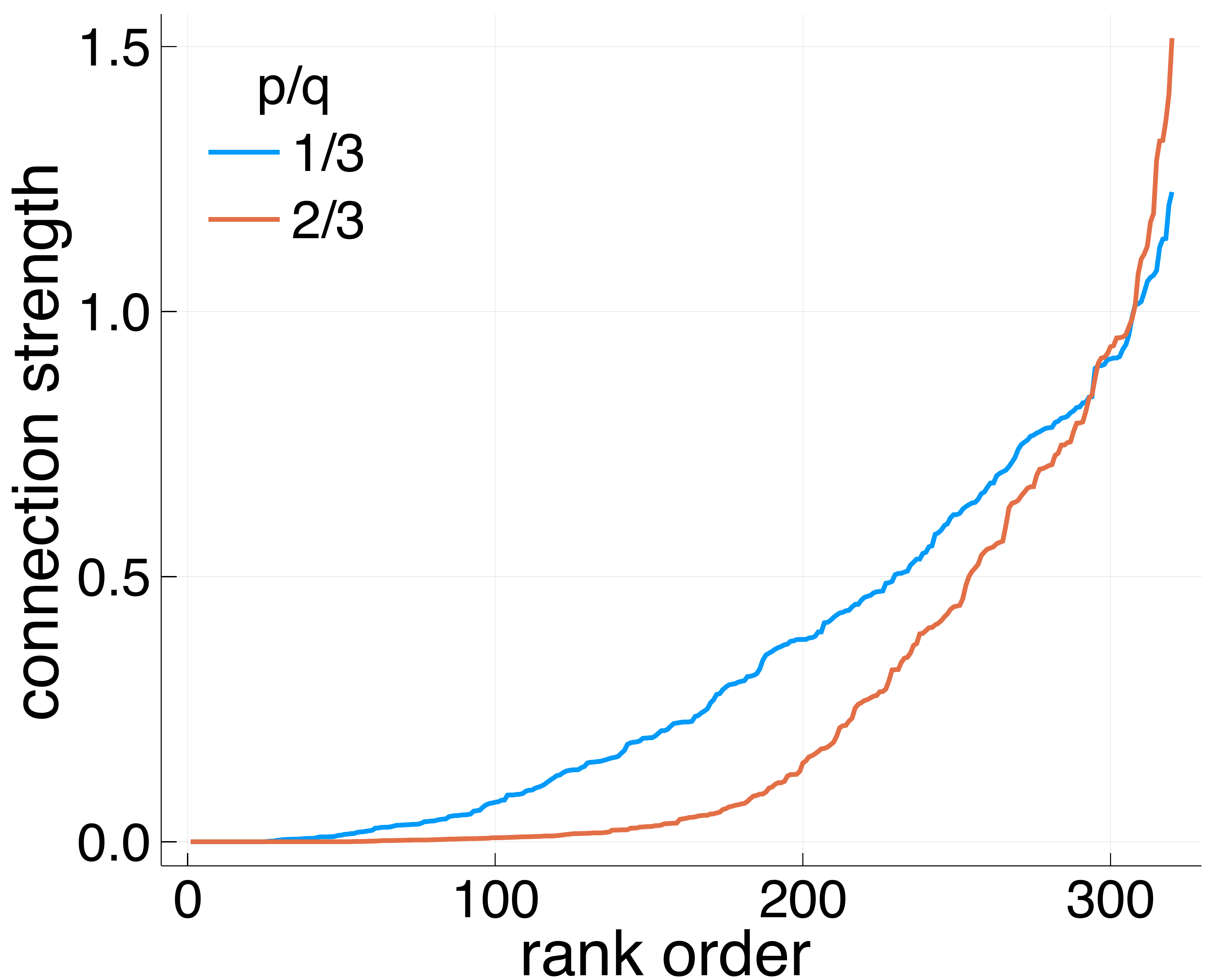}
\par\end{centering}
\caption{Sparser $E$ activity corresponds with a fuller $A$ matrix. The elements
of $A$ are graphed after sorting in increasing order. The matrix
is fuller for $p/q=1/3$ than for $p/q=2/3$. Activity of $E$ neurons
is sparser and more decorrelated for $p/q=1/3$, as was shown by Figs.
(\ref{fig:AllActivities}) and (\ref{fig:EECorrelationHistogram}).\label{fig:SparserActivityFullerA}}

\end{figure}

Visualization of the $E-I$ connections from numerical simulations
(Fig. \ref{fig:SparserActivityFullerA}) illustrates that a fuller
$A$ matrix corresponds to sparser $E$ activity. Small $p/q$ means
weaker competition and less sparse $A$ connections. At the same time,
it means sparser and more decorrelated $E$ activity, by Eq. (\ref{eq:CosineSimilarityBound}).

\begin{figure}
\begin{centering}
\includegraphics[width=0.8\textwidth]{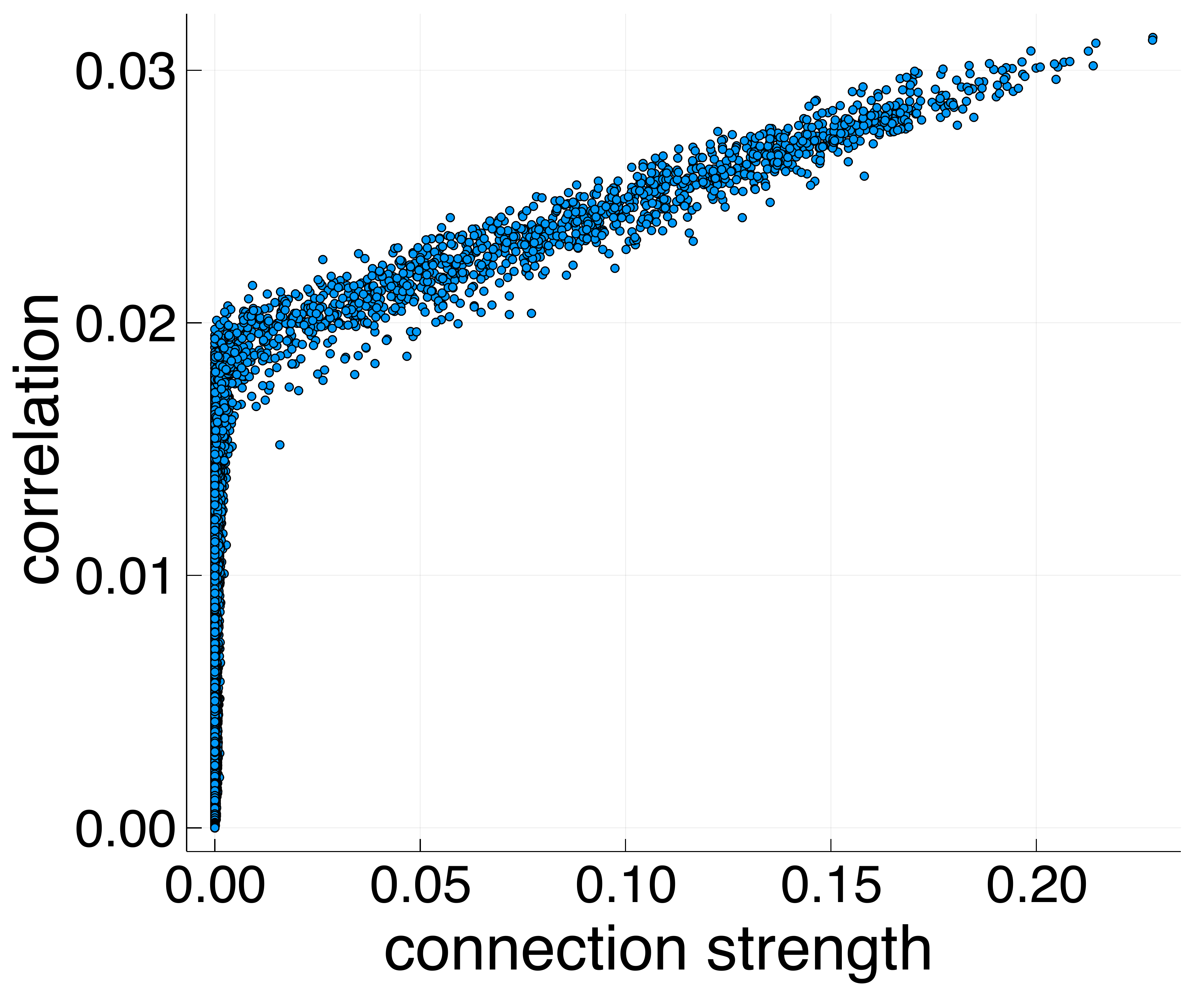}
\par\end{centering}
\caption{$S\to E$ connection strength is linearly related to correlation.
Correlation $\langle x_{i}u_{a}\rangle$ versus connection strength
$W_{ia}$ has the same slope and intercept for every $E$ neuron,
consistent with Eq. (\ref{eq:CorrelationConnectionFeedforwardExcitation})
and the fact that $\sum_{b}W_{ib}$ turns out to vary little across
$i$.\label{fig:ESCorrelationConnection}}

\end{figure}

The $S\to E$ connections can be analyzed similarly. Equation (\ref{eq:UpdateW})
and nonnegativity of $W$ imply that
\begin{equation}
\gamma W_{ia}=\left[\left\langle x_{i}u_{a}\right\rangle -\kappa\sum_{b}W_{ib}\right]^{+}\label{eq:CorrelationConnectionFeedforwardExcitation}
\end{equation}
at a stationary state of learning. Numerical simulations (Fig. \ref{fig:ESCorrelationConnection})
agree well with Eq. (\ref{eq:CorrelationConnectionFeedforwardExcitation}).
For each $E$ neuron $i$, the top $k$ connections are linearly related
to the top $k$ correlations, while the rest of the connections vanish.
The number $k$ of surviving connections varies from neuron to neuron,
and is governed by an inequality derived in Appendix \ref{sec:NumberSurvivingConnections}.

\section{Excitatory-inhibitory balance}

\begin{figure}
\begin{centering}
\includegraphics[width=1\textwidth]{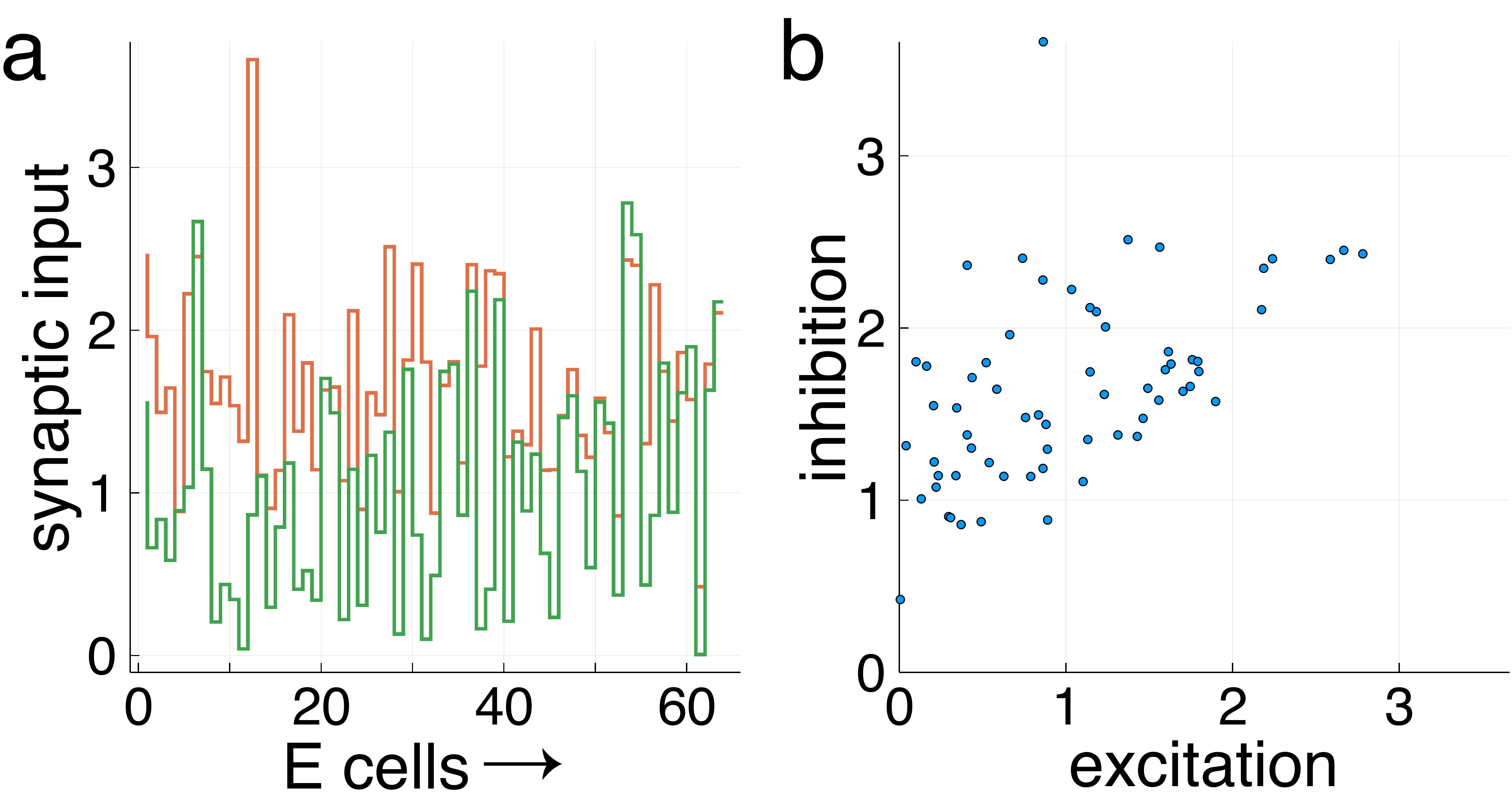}
\par\end{centering}
\caption{Excitatory-inhibitory balance as a byproduct of learning via Hebbian
excitation and anti-Hebbian inhibition. (a) Excitatory input $\lambda_{i}^{-1}\sum_{a}W_{ia}u_{a}$
(green) and inhibitory input $\lambda_{i}^{-1}\sum_{\alpha}y_{\alpha}A_{\alpha i}$
(red) to each $E$ cell $i$ for a single stimulus after convergence
of the activity dynamics. Only a few $E$ cells are active, those
for which excitation (green) exceeds inhibition (red). (b) Excitation
only slightly exceeds inhibition for active cells. Active $E$ cells
correspond to points below the main diagonal (excitation equals inhibition),
and these points are only slightly below.\label{fig:BalancedEI}}

\end{figure}

At a steady state of the activity dynamics, Eq. (\ref{eq:ExcitatoryDynamics})
implies that
\[
x_{i}=\left[\lambda_{i}^{-1}\left(\sum_{a=1}^{m}W_{ia}u_{a}-\sum_{\alpha=1}^{r}y_{\alpha}A_{\alpha i}\right)\right]^{+}
\]
The activity of the $i$th $E$ cell is determined by the difference
between its excitatory input $\lambda_{i}^{-1}\sum_{a}W_{ia}u_{a}$
(due to $S\to E$ connections) and its inhibitory input $\lambda_{i}^{-1}\sum_{\alpha}y_{\alpha}A_{\alpha i}$
(due to $I\to E$ connections). Active $E$ cells are those for which
excitatory input exceeds inhibitory input. Numerical simulations show
that excitatory input only slightly exceeds inhibitory input for active
cells (Fig. \ref{fig:BalancedEI}). This is reminiscent of a phenomenon
reported for cortical neurons and known as excitatory-inhibitory balance
\citep{isaacson2011inhibition}. Previous computational models of
balanced networks have focused on explaining Poisson-like variability
of spiking activity \citep{deneve2016efficient}. The current network
is not intended as an explanation of spiking variability, as it is
a rate-based model. The novelty is that excitatory-inhibitory balance
emerges from synaptic plasticity rather than the nonmodifiable random
or structured connectivity of previous models \citep{deneve2016efficient}.

\section{Discussion}

Numerical simulations have shown that a modest number of $I$ neurons
can be sufficient to decorrelate the activities of $E$ neurons, consistent
with the fact that inhibitory neurons are a small minority of cortical
neurons. Rather complete decorrelation is achievable without $I-I$
connections. Including $I-I$ connections, as in the prior spiking
network model of \citet{king2013inhibitory}, would add more neurobiological
realism as well as computational power. Future work will consider
the computational roles of $E-E$ as well as $I-I$ connections.

Many classes of inhibitory neurons are known to exist in the cortex.
\citet{king2013inhibitory} proposed that the anti-Hebbian $I$ neurons
in their model should be identified with parvalbumin-positive fast-spiking
basket cells, which are fast, have high average firing rates, are
reciprocally coupled with excitatory cells, and exhibit plasticity.
This class of cells is also a reasonable candidate for the $I$ neurons
in the present model.

The number of connections in a network with disynaptic inhibition
is much reduced relative to a network with all-to-all inhibition.
Therefore disynaptic inhibition can be regarded as a more efficient
way of achieving decorrelation. In this view, incompleteness of decorrelation
is something to be eliminated or reduced as much as possible. Alternatively,
one can imagine that incompleteness of decorrelation could have an
important computational function. Perhaps the remaining correlations
in the output of the network could serve as a basis for further learning
by another network.

\section*{Acknowledgments}

The author is grateful for helpful discussions with C. Pehlevan and
D. Chklovskii. The research is supported in part by the Intelligence
Advanced Research Projects Activity (IARPA) via DoI/IBC contract number
D16PC0005, and by the National Institutes of Health via U19 NS104648
and U01 NS090562.

\appendix

\section{Optimization by the gradient projection algorithm\label{sec:GradientProjection}}

Eq. (\ref{eq:ExcitatoryDynamics}) is a diagonally rescaled gradient
projection algorithm,

\begin{align}
\vec{x} & :=\left[\left(1-dt\right)\vec{x}-dt\,\Lambda^{-1}\partial L/\partial\vec{x}\right]^{+}\label{eq:GradientProjection}\\
 & :=\left[\left(1-dt\right)\vec{x}+dt\,\Lambda^{-1}\left(W\vec{u}-A^{\top}A\vec{x}\right)\right]^{+}\nonumber 
\end{align}
for solving the optimization problem 
\begin{equation}
\min_{\vec{x}\geq0}L(\vec{x})\label{eq:Optimization}
\end{equation}
where
\begin{align}
L(\vec{\vec{x}}) & =\frac{1}{2}\vec{x}^{\top}\left(\Lambda+A^{\top}A\right)\vec{x}-\vec{x}^{\top}W\vec{u}\label{eq:LyapunovFunction}\\
 & =\frac{1}{2}\sum_{i}\lambda_{i}x_{i}^{2}+\frac{1}{2}\sum_{\alpha}\left(\sum_{i}A_{\alpha i}x_{i}\right)^{2}-\sum_{ia}W_{ia}x_{i}u_{a}\nonumber 
\end{align}
A simple backtracking line search was found to be more efficient than
a fixed value of $dt$. Namely, if a gradient projection step causes
$L$ to increase, then the step is rejected and the step size parameter
$dt$ is halved. If $L$ does not increase, the step is accepted,
and $dt$ is increased by 1 percent with a ceiling of 0.5. The iteration
terminates when the root mean square of $\partial L/\partial x_{i}$
for $i$ such that $x_{i}>0$ is less than $10^{-3}$. For each stimulus,
$dt$ is initialized at 0.4.

\section{Number of surviving connections\label{sec:NumberSurvivingConnections}}

\subsection{Disynaptic inhibition}

Without loss of generality, assume that $\langle y_{\alpha}x_{1}\rangle\geq\langle y_{\alpha}x_{2}\rangle\geq\ldots\geq\langle y_{\alpha}x_{n}\rangle$.
Then we have $A_{\alpha1},\ldots,A_{\alpha k}>0$ and $A_{\alpha,k+1}=\ldots=A_{\alpha n}=0$
for $k$ satisfying 
\[
\langle y_{\alpha}x_{k+1}\rangle\leq p^{2}\sum_{i=1}^{k}A_{\alpha i}<\langle y_{\alpha}x_{k}\rangle
\]
Summing Eq. (\ref{eq:CorrelationConnectionDisynapticInhibition})
over $j=1,\ldots,k$ yields
\[
\left(q^{2}-p^{2}\right)\sum_{j=1}^{k}A_{\alpha j}=\sum_{j=1}^{k}\langle y_{\alpha}x_{j}\rangle-p^{2}k\sum_{i=1}^{k}A_{\alpha i}
\]
from which it follows that
\[
\sum_{j=1}^{k}A_{\alpha j}=\frac{1}{q^{2}+(k-1)p^{2}}\sum_{j=1}^{k}\langle y_{\alpha}x_{j}\rangle
\]
Substituting back into the first inequality yields
\[
\langle y_{\alpha}x_{k+1}\rangle\leq\frac{p^{2}}{q^{2}+(k-1)p^{2}}\sum_{j=1}^{k}\langle y_{\alpha}x_{j}\rangle<\langle y_{\alpha}x_{k}\rangle
\]
Note that this condition depends on $p$ and $q$ only through their
ratio $p/q$. If correlations are held fixed, the competition is ``winner-take-all''
($k=1$) as $p/q\to1$, with only one nonzero connection surviving.
The condition for all connections to survive is
\[
\frac{\sum_{j=1}^{n}\langle y_{\alpha}x_{j}\rangle}{\langle y_{\alpha}x_{n}\rangle}\leq\frac{q^{2}+(n-1)p^{2}}{p^{2}}
\]
which is satisfied as $p/q\to0$ if correlations are held fixed.

\subsection{Feedforward excitation}

Without loss of generality, assume that $\langle x_{i}u_{1}\rangle\geq\langle x_{1}u_{2}\rangle\geq\ldots\geq\langle x_{1}u_{m}\rangle$.
Then we have $W_{i1},\ldots,W_{ik}>0$ and $W_{i,k+1}=\ldots=W_{im}=0$
for $k$ satisfying 
\[
\langle x_{i}u_{k+1}\rangle\leq\frac{1}{\gamma/\kappa+k}\sum_{a=1}^{k}\langle x_{i}u_{a}\rangle<\langle x_{i}u_{k}\rangle
\]
The competition is winner-take-all as $\gamma/\kappa\to0$.

\bibliographystyle{plainnat}
\bibliography{Foldiak}

\end{document}